\documentclass[twocolumn,prl]{revtex4}
\usepackage{graphicx}
\usepackage{latexsym}

\begin{document}

\title{Critical Resistance of the Quantum Hall Ferromagnet in AlAs 2D Electrons}
\author{E. P. De Poortere, E. Tutuc, and M. Shayegan}
\address{Department of Electrical Engineering, Princeton University, Princeton, New Jersey 08544}
\date{\today}

\begin{abstract}
Magnetic transitions in AlAs two-dimensional electrons give rise to sharp resistance spikes within the quantum
Hall effect. Such spikes are likely caused by carrier scattering at magnetic domain walls below the Curie
temperature. We report a critical behavior in the temperature dependence of the spike width and amplitude, from
which we deduce the Curie temperature of the quantum Hall ferromagnet. Our data also reveal that the Curie
temperature increases monotonically with carrier density.
\end{abstract}


\maketitle

\begin{figure*}
     \centering
     \includegraphics[scale=0.65]{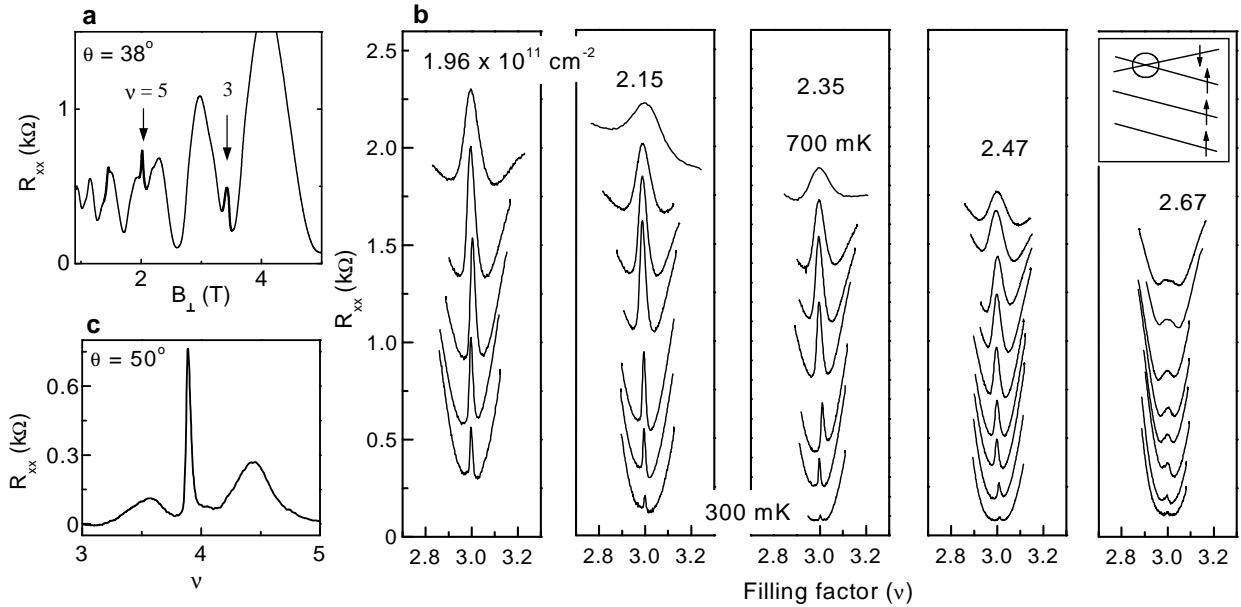}
     \caption{(a) Magnetoresistance of AlAs 2D electrons for a density $n = 2.5 \times 10^{11}$ cm$^{-2}$. Resistance
     spikes occur at Landau level crossings near $\nu = 3$ and 5. (b) Magnetoresistance of AlAs 2D electrons near $\nu = 3$, showing the resistance spike for different $n$ and temperatures (300 $\lesssim T \lesssim$ 700 mK). From left to right, tilt angles are $\theta
= 16^o$, $25^o$, $30^o$, $34^o$, and $37^o$. Note the strong suppression of the spike amplitude as $n$ increases.
For clarity, only data for downward $B$ sweeps are shown. Inset schematically shows the crossing of opposite spin
Landau levels at $\nu = 3$. (c) Resistance spike near $\nu = 4$ for $n = 1.86 \times 10^{11}$ cm$^{-2}$.}
    \label{allspikes}
\end{figure*}

Various forms of order have been shown to derive from or compete with the quantum Hall (QH) state in
two-dimensional (2D) electron systems, including charge density waves, the Wigner crystal, stripe and bubble
phases, and liquid crystal states. Recent experimental evidence, often in the form of magnetic hysteresis, also
points to the existence of a ferromagnetic phase within the integer and fractional QH states
\cite{koch93,daneshvar97,jungwirth98,cho98,piazza99,eom00,depoortere00,smet01,muraki01,jaroszynski02,chokomakoua03}.
The analogy with spontaneous polarization in conventional Ising ferromagnets exists when two Landau levels (LL's)
with opposite spin quantum numbers and different orbital indices are degenerate [Fig.\ \ref{allspikes}(b), inset].
Experimentally, this LL coincidence is obtained by tilting the normal of the sample at an angle $\theta$ with
respect to the magnetic field. Exchange energy ensures that at temperature $T = 0$, one of the two crossing levels
is fully occupied and the other empty. Magnetic domains form as the temperature is raised, and the electronic
state within each domain is described as an Ising-like QH ferromagnet with either one of two possible spin
orientations. In materials such as AlAs \cite{depoortere00}, (Cd,Mn)Te \cite{jaroszynski02}, InGaAs \cite{koch93},
InSb \cite{chokomakoua03}, and wide GaAs quantum wells \cite{muraki01}, LL crossings and the associated magnetic
domains lead to resistance maxima within the integer QH or Shubnikov-de Haas minima of the magnetoresistance
($R_{xx}$) (Fig.\ \ref{allspikes}), while in GaAs 2D electrons, sharp resistance peaks are observed in the
fractional QH $R_{xx}$ minima. In the latter case, the spikes are caused by crossing LL's of composite fermions
\cite{smet01}.

\begin{figure}
      \centering
      \includegraphics[scale=0.4]{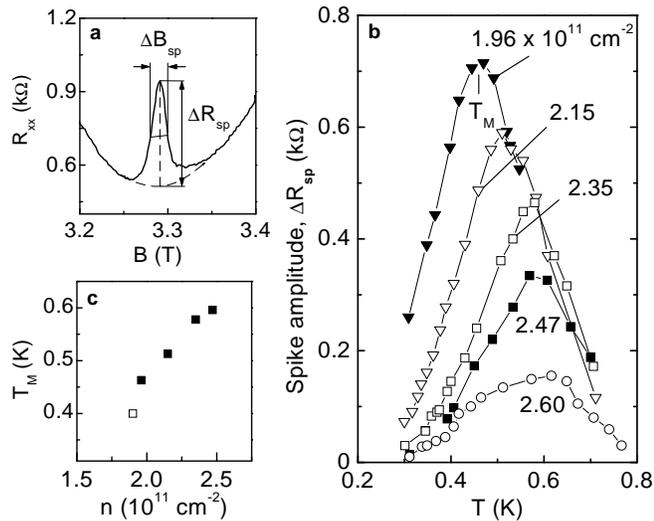}
      \caption{(a) Definition of the spike amplitude ($\Delta R_{sp}$) and linewidth ($\Delta B_{sp}$). (b) Plots of the resistance spike amplitude ($\Delta R_{sp}$) vs. $T$, for several carrier densities. (c) Temperatures of the $\Delta R_{sp}$ maxima as a function of $n$ (full squares: from this figure; hollow square: data from a separate cooldown).}
    \label{spikeampl}
\end{figure}

Theorists have begun to address the nature of domain boundaries in Ising QH ferromagnets
\cite{jungwirth01b,brey02,chalker02}, though no calculation of the domain wall resistance in these systems exists
at present. Estimates for the domain wall energies, however, yield the Curie temperature of the ferromagnet
\cite{jungwirth01b}, which can be determined experimentally. Determination of this temperature is the first
objective of our work. Furthermore, AlAs resistance spikes can be separated fairly easily from the background
resistance of the 2D system, so that they provide a direct measure of the domain wall magnetoresistance.
Measurements of the resistance spike dimensions, namely its amplitude ($\Delta R_{sp}$) and width in magnetic
field ($\Delta B_{sp}$) [see Fig.\ \ref{spikeampl}(a)], therefore supply an experimental basis against which
future models of the domain walls can be tested, and provide a second motivation for our study.

In this Letter, we report temperature- and density-dependence measurements of the resistance spikes in AlAs 2D
electrons. Our results reveal that: (1) $\Delta R_{sp}$ increases with $T$ up to a characteristic temperature,
$T_M$, beyond which it decreases with $T$; (2) below the same temperature $T_M$, $\Delta B_{sp}$ is relatively
constant, while above $T_M$, $\Delta B_{sp}$ increases significantly with temperature; (3) $T_M$ increases with
the 2D electron density ($n$). Comparing the temperature dependence of the spike width with that expected from a
mean-field 2D Ising model, we are able to identify $T_M$ with the Curie temperature of the QH ferromagnet.

AlAs offers several advantages for the study of QH ferromagnets: first, the band $g$-factor of 2D electrons in
AlAs is approximately 2 \cite{vankesteren89}, while the effective electron masses (at the $X$ points of the
Brillouin zone) are 1.1 and 0.19 in the longitudinal and transverse directions, respectively \cite{adachi85}. The
latter properties ensure that the Zeeman ($E_Z$) and cyclotron ($E_c$) energies are comparable in magnitude for
low values of $\theta$, a regime in which the orbital effect of the titled field can be neglected
\cite{jungwirth01b}. Second, the high quality of our AlAs quantum wells (QW's), in which we have observed
developing fractional QH states at high-order fillings \cite{depoortere02a}, implies that the interparticle
Coulomb interaction, rather than disorder, dominates the ground state of the 2D system.

The sample studied here is a 150 \AA-wide AlAs QW surrounded by Al$_{0.4}$Ga$_{0.6}$As barriers, modulation-doped
on the front side with Si, and grown by molecular beam epitaxy on a GaAs (411)B substrate. For magnetotransport
measurements, a $100 \times 325 \mu$m Hall bar was lithographically etched on the sample surface, and 1500
\AA-thick AuGeNi contacts were deposited and alloyed at 440 $^o$C for 10 minutes in forming gas. Front and back
gates were fitted to the sample. Experiments were performed in pumped $^3$He and dilution refrigerators in
magnetic fields up to 18 T. Prior to measurements, the sample was illuminated with a red LED at $T \simeq 4$ K,
while a positive bias was applied between the back gate and the 2D system. This procedure is required to make
Ohmic contact to the 2D system in our sample, and is detailed in \cite{depoortere03}.

Figure \ref{allspikes}(a) gives an overview of the sample magnetoresistance for $n = 2.5 \times 10^{11}$ cm$^{-2}$
and $\theta = 38^o$, showing the resistance spikes near $\nu = 3$ and 5. The $T$ and $n$ dependence of the $\nu =
3$ spike are given in Fig.\ \ref{allspikes}(b) \cite{constantnu}. From these data we first plot, in Fig.\
\ref{spikeampl}(b), the spike amplitudes as a function of $T$. We see that $\Delta R_{sp}$ reaches a maximum at a
temperature $T_M$, and that $T_M$ increases monotonically with $n$ [Fig.\ \ref{spikeampl}(c)]. Furthermore,
$\Delta R_{sp}$ decreases quickly as the {\it density} increases; for $n \gtrsim 2.7 \times 10^{11}$ cm$^{-2}$ in
this sample, the spike amplitude is too small to be measured. On the other hand, at the lowest measured density,
the maximum $\Delta R_{sp}$ is about 700 $\Omega$, comparable to the $R_{xx}$ maxima at half-integer $\nu$, when
the Fermi level lies close to the energies of extended states. We note that an even stronger resistance spike is
seen near $\nu = 4$ in this sample: its $\Delta R_{sp}$ is more than twice the magnitude of the neighboring
$R_{xx}$ maxima at or near half-integral fillings [Fig.\ \ref{allspikes}(c)]. The large magnitude of $\Delta
R_{sp}$ at low $n$ and near $\nu = 4$ indicates that the scattering process at the resistance spike is
fundamentally different from that present in the sample when the Fermi level nearly coincides with that of an
extended state. In the following paragraphs, we concentrate on the resistance spike near $\nu = 3$.

\begin{figure}
      \centering
      \includegraphics[scale=0.6]{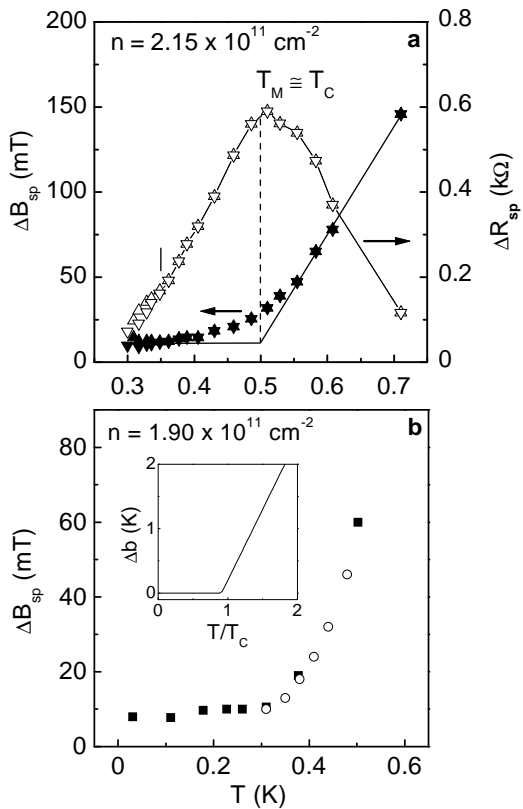}
      \caption{(a) $\Delta R_{sp}$ and $\Delta B_{sp}$ vs. $T$, showing the correspondence between the threshold temperature
      in $\Delta B_{sp}(T)$ with $T_M$, the temperature of the $\Delta R_{sp}$ maximum. $T_M$ is interpreted as the Curie temperature of
the QH ferromagnet ($T_C$). Upward- and downward-pointing triangles represent data from up and down field sweeps
respectively, and show that the onset of hysteresis (vertical mark at $T = 350$ mK) is significantly lower than
$T_M \simeq T_C$. (b) $\Delta B_{sp}(T)$ for $n = 1.90 \times 10^{11}$ cm$^{-2}$, showing $\Delta B_{sp}$
independent of $T$ at low temperatures, and increasing with $T$ for $T \gtrsim 400$ mK. Full and hollow symbols
are used to distinguish data from two different cooldowns. Inset: spike width predicted by Bragg-Williams theory,
displaying a behavior qualitatively similar to that of the measured $\Delta B_{sp}(T)$. $T_C$ in this graph is the
Curie temperature of the Ising ferromagnet.}
    \label{tempdep}
\end{figure}

The $T$ dependence of the spike width, $\Delta B_{sp}$, is plotted for two different densities in Figs.\
\ref{tempdep}(a) and (b). Both graphs show that $\Delta B_{sp}$ decreases as the 2D electrons cool down, and tends
to saturate below a given temperature that depends on the density. Although $\Delta B_{sp}$ depends smoothly on
$T$, the onset of its rise with $T$ appears to occur near $T_M$, the temperature at which $\Delta R_{sp}$ is
maximum. More quantitatively, a ``threshold'' temperature can be defined for the $\Delta B_{sp}(T)$ data in Figs.\
\ref{tempdep}(a) and (b), by first fitting straight lines to the low- and high-$T$ ranges, and taking the
intersection of these lines. In Fig.\ \ref{tempdep}(a), the resulting $T$ is 500 mK, a value close to $T_M$. This
correspondence suggests a common physical process for the $T$ dependencies of both $\Delta R_{sp}$ and $\Delta
B_{sp}$.

In order to better understand our data, we use a Bragg-Williams model \cite{chaikin95} applied to a system
composed of two energy levels with respective relative fillings $f_1$ and $f_2$, corresponding to the highest
occupied level and the lowest unoccupied level at $\nu = 3$. We define the magnetization ($m_z$) as $m_z = f_1 -
f_2$, so that $m_z = 1$ (-1) if the first (second) level is fully occupied. The electron energy can then be
written as $E(m_z) = bm_z - \frac{1}{2}Jm^2_z$, where $b$ is a reduced magnetic field ($b = 0$ at the transition)
that includes contributions from single-particle energies ($E_Z$ and $E_c$) and from interactions, and $J$ is an
effective interaction strength \cite{jungwirth01a}. The free energy of the 2D system is given by
\begin{equation} F(b,m_z) = -TS(m_z) + E(b,m_z), \label{eq:freeenergy}
\end{equation} \noindent
where $S(m_z)$ is the mixing entropy of the two phases \cite{chaikin95}. For every $b$ and $T$, we minimize $F$
with respect to $m_z$, and thus obtain the equilibrium magnetization. The model predicts that for $T < T_C = J$
(the Curie temperature), two different $m_z$'s minimize $F$, and that the system divides into magnetic domains.

Above the Curie point, the transition in field becomes increasingly gradual as $T$ increases. The width of the
transition can be quantified by the reduced field ($b_{+}$) at which $m_z = -1/2$, halfway between the transition
($m_z = 0$) and full polarization ($m_z = -1$). By symmetry, the field width of the transition, defined between
$m_z = +1/2$ and -1/2, is thus $\Delta b = 2b_{+}$. From Eq.\ \ref{eq:freeenergy} we derive:
\begin{equation}
\Delta b = 2b_{+}(T) = -T_C + 1.1T \hspace{2cm} (T > T_C). \label{eq:spikewidthT}
\end{equation} \noindent
We see that $\Delta b$, plotted in the inset of Fig.\ \ref{tempdep}(b) as a function of $T$, vanishes at $T$ close
to $T_C$ (i.e., at $T = 0.91 T_C$), and increases for larger $T$. This behavior is similar to that of the measured
$\Delta B_{sp}$, suggesting that the ``threshold'' temperatures in Figs. \ref{tempdep}(a) and (b) are close to the
Curie temperatures of the QH ferromagnet at the respective carrier densities. An implication of our analysis is
that $T_M \simeq T_C$, since the peak in $\Delta R_{sp}(T)$ agrees with the temperature threshold of $\Delta
B_{sp}(T)$.

The fact that the $\Delta R_{sp}$ maximum occurs near $T_C$ is qualitatively consistent with calculations of the
critical resistance in ferromagnetic semiconductors \cite{degennes58,alexander76} and with recent experiments in
(Ga,Mn)As \cite{ohno01}. According to those calculations, the diverging correlation length of spin fluctuations as
$T \rightarrow T_C$ causes a peak resistance at a temperature $T_M$ close to $T_C$, with $|T_M-T_C|/T_C \sim
10^{-5}$ in typical semiconductors \cite{alexander76}. Although the origin of scattering at the spike in our
system is unclear at present, data in Fig.\ \ref{spikeampl}(b) confirm qualitatively the predicted peak resistance
near $T_C$. We also note that we obtain similar empirical correspondences between the $T$ dependencies of $\Delta
R_{sp}$ and $\Delta B_{sp}$ for $n = 2.35$ and $2.47 \times 10^{11}$ cm$^{-2}$, two other densities for which the
data ranges are large enough for comparison to be possible.

In Fig.\ \ref{spikeampl}(c), we plot the density dependence of $T_M$ deduced from the $\Delta R_{sp}$ peaks in
Fig.\ \ref{spikeampl}(b). Within the small density range allowed by our experiment, we observe that $T_M$
increases with $n$. This is consistent with our expectation that $T_C \simeq T_M$, through the exchange
interaction, should scale with the Coulomb energy, $e^2/4\pi \epsilon l_B$ [$\epsilon \simeq 10 \epsilon_0$ is the
dielectric constant of AlAs and $l_B = (\hbar/eB_{\bot})^{1/2}$ is the magnetic length] \cite{macdonald86}: the
latter increases with $B_{\bot}$, which in turn increases with $n$ at a fixed $\nu = 3$. The values we obtain for
$T_C$ also agree with the estimate ($\sim 500$ mK) calculated by Jungwirth and MacDonald for the Curie temperature
of AlAs 2D electrons at $n = 2.5 \times 10^{11}$ cm$^{-2}$ \cite{jungwirth01b}.

Plotted in Fig.\ \ref{tempdep}(a) are the $\Delta R_{sp}$ and $\Delta B_{sp}$ data for both upward and downward
$B$ sweeps. These values, which depend on the sweep direction at sufficiently low $T$, reflect the observed
$R_{xx}$ hysteresis (which mostly occurs in amplitude rather than in field). Figure \ref{tempdep}(a) shows that
the temperature corresponding to the onset of hysteresis is {\it lower} than the Curie temperature derived from
the $\Delta R_{sp}(T)$ and $\Delta B_{sp}(T)$ dependencies. This implies that magnetic domains present (below
$T_C$) at the LL crossing do not necessarily give rise to hysteretic $R_{xx}$. Furthermore, the strength of
hysteresis depends on sample cooldown, while $T_C$ as defined above does not show such sensitivity, suggesting
that hysteresis is controlled by cooldown-dependent parameters such as the precise nature of impurity disorder.

While the $T$ and $n$ dependencies described in our work apply to resistance spikes at $\nu = 3$, we have obtained
similar data for LL crossings at other filling factors, though in a less complete manner. The $T$ dependencies of
$\Delta R_{sp}$ and $\Delta B_{sp}$ for the spike near $\nu = 5$, e.g., are analogous to our $\nu = 3$ spike data;
the amplitude of the spike at $\nu = 4$ (at $T = 30$ and 300 mK) also decreases monotonically as $n$ increases.
Spike measurements at $\nu = 4$ and 5 are thus qualitatively consistent with results obtained from the $\nu = 3$
spike.

We now outline the main differences between our work and that of Jaroszy\'{n}ski {\it et al.}
\cite{jaroszynski02}, who recently performed a detailed study of resistance spikes in (Cd,Mn)Te quantum wells.
First, authors in \cite{jaroszynski02} determine $T_C$ from the temperature at which the {\it total} value of the
resistance at the spike ($R_{tot}$) reaches a maximum, whereas we treat the spike {\it amplitude} $\Delta R_{sp}$
as the physical parameter reflecting the contribution of the domains (or, near $T_C$, of the spin fluctuations) to
the total resistance. Second, the onset of hysteresis in Ref.\ \cite{jaroszynski02} matches the peak in
$R_{tot}(T)$ (for one of the measured densities). In our samples, as can be seen in Fig.\ \ref{allspikes}(b),
$R_{tot}$ increases monotonically with $T$, so that the peak in $R_{tot}(T)$, if it exists, occurs at $T > 700$
mK, i.e., at a much higher $T$ than the onset of spike hysteresis \cite{hyst}. Third, $T_C$ determined in Ref.\
\cite{jaroszynski02} {\it decreases} with increasing $n$, a trend that is contrary to our experimental results and
unexpected based on theoretical grounds. While we do not understand the origin of these various differences
between spikes in AlAs and (Cd,Mn)Te, we emphasize that the latter contains magnetic Mn ions that may play a role
in the formation of magnetic domains, while AlAs is nonmagnetic.

In conclusion, we have observed critical behaviors (at matching temperatures) in both the amplitude and the
linewidth of the resistance spikes in the AlAs QH ferromagnet, and determined from these the Curie temperature of
the ferromagnet. The critical behaviors are qualitatively consistent with the results of two independent models:
the observed peak in $\Delta R_{sp}(T)$ is predicted by calculations of the critical resistance in ferromagnetic
semiconductors, and the measured $\Delta B_{sp}(T)$ dependence can be understood by a mean-field analysis of the
Ising model.


This work was supported by the Princeton University NSF MRSEC grant. We acknowledge fruitful discussions with
Ravin Bhatt, Duncan Haldane, and Allan MacDonald, and thank Eric Palm and Tim Murphy for their technical
assistance. Part of this work was performed at the National High Magnetic Field Laboratory in Tallahassee, FL,
which is also supported by the NSF.


\begin{references}
\bibitem{constantnu}
For each $n$ and $T$ in Fig.\ \ref{allspikes}(b), the tilt angle was chosen so that the transition (i.e., the
resistance spike) would occur as close to $\nu = 3$ as possible. This angle turning is necessary because for a
fixed tilt angle, the spike shifts to slightly lower $B$ as $T$ increases. In practice, the angle tuning we
perform to keep the spike at $\nu = 3$ results in a 3 \% shift of the spike field. To first order, the spike field
can be considered constant.
\end{references}
\end{document}